\documentclass{emulateapj}
\shorttitle{Millimeter Imaging of HH 24 MMS}
\shortauthors{Kang et al.}

\begin{document}

\title{Millimeter Imaging of HH 24 MMS: A Misaligned Protobinary System}
\author{\sc Miju Kang\altaffilmark{1,2,3},
            Minho Choi\altaffilmark{1},
            Paul T. P. Ho\altaffilmark{4,5},
            and Youngung Lee\altaffilmark{1}}
\altaffiltext{1}{International Center for Astrophysics,
                 Korea Astronomy and Space Science Institute,
                 Hwaam 61-1, Yuseong, Daejeon 305-348, South Korea.;
                 mjkang@kasi.re.kr.}
\altaffiltext{2}{Department of Astronomy and Space Science,
                 Chungnam National University, Daejeon 305-764, South Korea.}
\altaffiltext{3}{Steward Observatory, 
				 University of Arizona, 
				 933 North Cherry Avenue, Tucson, AZ 85721.}
\altaffiltext{4}{Academia Sinica Institute of Astronomy and Astrophysics,
                 Taipei 106, Taiwan.}
\altaffiltext{5}{Harvard-Smithsonian Center for Astrophysics,
                 Cambridge, MA 02138.}
\setcounter{footnote}{5}

\begin{abstract}
The HH 24 MMS protostellar system was observed in the 6.9 mm continuum
with a high angular resolution (0.5$''$). 
HH 24 MMS was resolved into two sources.
The separation between sources 1 and 2 is $\sim$0.9$''$ or 360 AU.
The spectral energy distribution 
suggests that the 6.9 mm flux is almost entirely from dust.
The 6.9 mm image and the spectrum suggest that 
HH 24 MMS may be a protostellar binary system. 
Total mass including the accretion disks and the inner protostellar envelope
is $\sim$1.4 $M_\odot$. 
Disk masses of sources 1 and 2 are 0.6 and 0.3 $M_\odot$, respectively.
Both sources are highly elongated.
The difference in the position angle of the two disks is $\sim$45\arcdeg, 
which means that HH 24 MMS is a highly misaligned protobinary system.
The misalignment suggests that turbulent fragmentation
may be the formation mechanism relevant to the binary systems
with a separation of a few hundreds of AU, such as the HH 24 MMS system.
\end{abstract}


\section{INTRODUCTION}

Young stellar objects are often classified
using the spectral energy distribution
(SED; Lada 1987; Andr{\'e} et al. 1993).
Class 0 sources have cold spectra 
that peak in the submillimeter/far-IR band.
Class 0 sources are interpreted as highly embedded protostars
still in their main accretion phase (Andr{\'e} \& Montmerle 1994). 
A protostar is thought to contain a star-disk system 
surrounded by a collapsing envelope (e.g., Shu et al. 1987).
It is important to understand the structure of Class 0 objects 
since they represent the earliest phase of star formation. 
However, it is not easy to study such systems owing to the high 
extinction in the optical or IR wavelength shorter than $\sim$20 $\mu$m.
Observations in the millimeter band provide us 
images of protostellar envelopes and deeply embedded accretion disks.  
High-resolution millimeter images often show
that many Class 0 sources are multiple protostellar systems
(Looney et al. 2000).

Many stars form as members of binaries or multiple systems
(e.g., Duquennoy \& Mayor 1991).
The multiplicity of young pre-main-sequence stars
is higher than or comparable to that of main-sequence stars
(Duch{\^e}ne et al. 2007),
implying that multiple systems form through fragmentation of cores
in the early stage of star formation.
Several kinds of fragmentation mechanisms have been suggested,
including rotational fragmentation, turbulent fragmentation,
and disk fragmentation (Goodwin et al. 2007).
Fragmentation of cloud cores has been investigated
through numerical simulations in various situations
including many kinds of physical and chemical effects,
and it was found that most cores tend to fragment into multiple objects
(see Goodwin et al. 2007 and references therein).
While collapsing cores fragment easily,
the relative importance of rotation and turbulence
in star forming regions remains unresolved.
Observations of Class 0 sources in the millimeter band
can provide clues to the fragmentation mechanism.

HH 24 MMS is a Class 0 source
located $\sim$20$''$ southwest of the Herbig-Haro object HH 24A
in the L1630 cloud
(Chini et al. 1993; Ward-Thompson et al. 1995; Bontemps et al. 1996).
It was detected in the 3.6 cm and 2.0 cm continuum
by Bontemps et al. (1996).
Chandler et al. (1995) suggested
that the dust emission of HH 24 MMS originates from two components:
an unresolved disk and an extended envelope. 
Bontemps et al. (1995) suggested
that the 3.6 cm emission at the position of the submillimeter peak
originated from thermal free-free emission,
but later they found
that the unresolved component of the centimeter continuum source
has an SED consistent with the dust disk (Bontemps et al. 1996).
Reipurth et al. (2002) imaged the 3.6 cm source
with an angular resolution of $\sim$1$''$.
High-density molecular gas was detected in molecular lines
(Gibb \& Heaton 1993).
Non-detection of X-rays from HH 24 MMS suggests
that either HH 24 MMS may be very deeply embedded
or HH 24 MMS does not have (proto)stellar flares 
to produce X-ray emission (Ozawa et al. 1999; Simon et al. 2004).

Despite the name, most Herbig-Haro objects in the HH 24 complex
are unrelated to HH 24 MMS
but belong to outflows driven by SSV 63 and its companions
(Mundt et al. 1991; Davis et al. 1997).
Near-IR observations revealed
that HH 24 MMS is driving its own outflow or an ``H$_2$ jet''
oriented in the northeast-southwest direction
(Bontemps et al. 1996; Davis et al. 1997).
Eisl{\"o}ffel \& Mundt (1997) suggested
that HH 24A and HH 24L may belong to the HH 24 MMS jet.
This jet was also detected in the radio continuum and high-velocity CO gas
(Bontemps et al. 1996).

\begin{deluxetable*}{lccccrcc}
\tabletypesize{\scriptsize}
\tablecaption{Continuum Sources in the HH 24 MMS Region}
\tablewidth{0pt}
\tablehead{
& \multicolumn{2}{c}{\sc Peak Position} & \colhead{\sc Peak Intensity}
& \multicolumn{2}{c}{\sc Source Size\tablenotemark{a}}
& \colhead{\sc Total Flux\tablenotemark{a}} & \colhead{\sc Mass} \\
\cline{2-3} \cline{5-6}
\colhead{\sc Source} & \colhead{$\alpha_{2000}$}
& \colhead{$\delta_{2000}$} & \colhead{(mJy beam$^{-1}$)}
& \colhead{FWHM} & \colhead{P.A.} & \colhead{(mJy)} & \colhead{($M_\odot$)}}
\startdata
1\dotfill & 05 46 08.37 & --00 10 43.7 & 2.54 $\pm$ 0.07
          & 0.72 $\pm$ 0.04 $\times$ 0.28 $\pm$ 0.03 & $-$66 $\pm$ 3
          & 4.9 $\pm$ 0.4 & 0.6$^{+0.6}_{-0.3}$ \\
2\dotfill & 05 46 08.42 & --00 10 43.1 & 0.87 $\pm$ 0.07
          & 1.35 $\pm$ 0.23 $\times$ 0.18 $\pm$ 0.05 &    69 $\pm$ 2
          & 3.0 $\pm$ 0.6 & 0.3$^{+0.4}_{-0.1}$ \\
3\dotfill & 05 46 08.56 & --00 10 42.9 & 0.45 $\pm$ 0.07
          & unresolved & \nodata & \nodata & \nodata \\
\enddata
\tablecomments{Units of right ascension are hours, minutes, and seconds,
               and units of declination are degrees, arcminutes,
               and arcseconds.
               Source parameters are
               from the C-array uniform-weight map (Fig. 1$b$).
               Fluxes were corrected for the primary beam response.}
\tablenotetext{a}{Parameters of the best-fit
                  two-component elliptical Gaussian profile.
                  Deconvolved size is in arcseconds,
                  and position angle is in degrees.
                  At 400 pc, the deconvolved sizes of sources 1 and 2
                  correspond to 290 AU $\times$ 110 AU
                  and 540 AU $\times$ 70 AU, respectively.}
\end{deluxetable*}

In this paper, we present high resolution images
of the Class 0 protostar HH 24 MMS observed in the 6.9 mm continuum
using the Very Large Array (VLA).
In \S~2 we describe our observations.
In \S~3 we report the results of the 6.9 mm imaging.
In \S~4 we discuss the physical properties of the dust continuum sources
in the HH 24 MMS region.

\setcounter{footnote}{5}

\section{OBSERVATIONS}
The HH 24 MMS region was observed
using the VLA of the National Radio Astronomy Observatory%
\footnote{The NRAO is a facility of the National Science Foundation
operated under cooperative agreement by Associated Universities, Inc.}
in the standard $Q$-band continuum mode (43.3 GHz or $\lambda$ = 6.9 mm)
in two observing tracks.
The phase and amplitude were determined
by observing the nearby quasar 0532+075 (PMN J0532+0732).
The phase tracking center was
$\alpha_{2000}$ = 05$^{\rm h}$46$^{\rm m}$08.38$^{\rm s}$ and
$\delta_{2000}$ = $-$00\arcdeg10$'$43.36$''$.
The first track was carried out in the D-array configuration
with twenty-five antennas on 2003 March 30.
The flux density of 0532+075 was bootstrapped
from the quasar 0542+498 (3C 147)
assuming that its flux density is 0.72 Jy,
which is the flux density measured within a day of our observations
(VLA Calibrator Flux Density Database%
\footnote{See http://aips2.nrao.edu/vla/calflux.html.}).
The bootstrapped flux density of 0532+075 is 1.05 Jy.
The second track was in the C-array configuration
with twenty-three antennas on 2004 March 2.
The flux calibration was done
by observing the quasar 0713+438 (QSO B0710+439).
The flux density of 0713+438 was set to 0.20 Jy,
which is the flux measured within 2 days of our observations
according to the VLA Calibrator Flux Density Database,
and the bootstrapped flux density of 0532+075 is 0.70 Jy.
Maps were made using a CLEAN algorithm.

\section{RESULTS}

\begin{figure*}
\epsscale{1.1}
\plotone{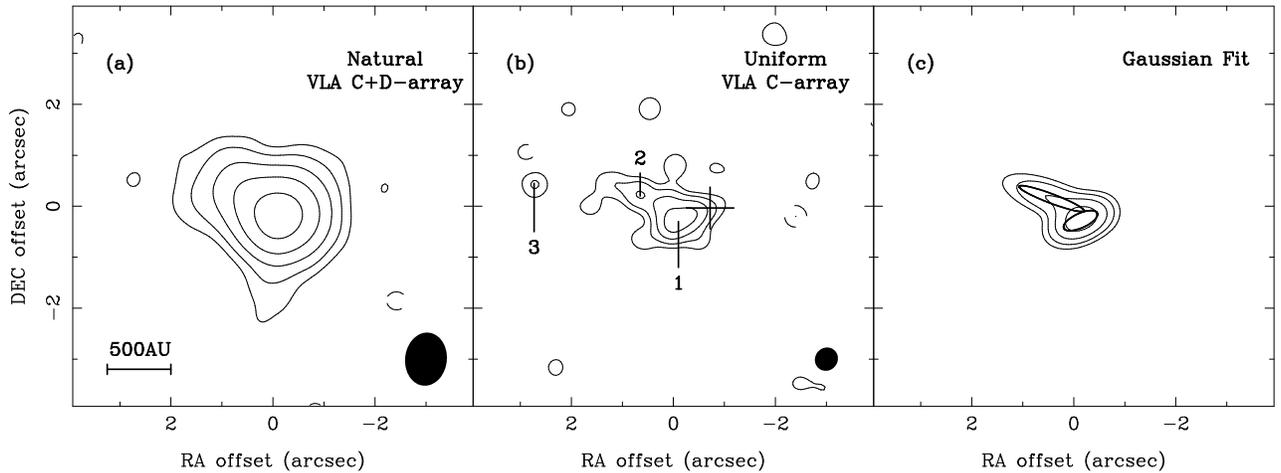}
\caption{
Maps of the $\lambda$ = 6.9 mm continuum toward the HH 24 MMS region.
($a$)
Natural-weight CLEAN map from the C- and the D-array data combined.
The contour levels are 1, 2, 4, 8, and 16 times 0.24 mJy beam$^{-1}$.
The rms noise is 0.08 mJy beam$^{-1}$.
Dashed contours represent negative levels.
Shown at the bottom right corner is the synthesized beam:
FWHM = 1.0$''$ $\times$ 0.8$''$ and P.A. = --6\arcdeg.
The straight line at the bottom left corner
corresponds to 500 AU at a distance of 400 pc.
($b$)
Uniform-weight CLEAN map from the C-array data only.
The contour levels are 1, 2, 4, and 8 times 0.21 mJy beam$^{-1}$.
The rms noise is 0.07 mJy beam$^{-1}$.
Source numbers are labeled.
The synthesized beam has
FWHM = 0.5$''$ $\times$ 0.4$''$ and P.A. = --39\arcdeg.
{\it Plus sign}:
Position of the 3.6 cm continuum source (Reipurth et al. 2002).
The size of the marker corresponds to the synthesized beam.
($c$)
Best-fit image of the complex containing sources 1 and 2
from the two-component elliptical Gaussian fit
to the map shown in panel ($b$).
{\it Thick ellipses}:
Deconvolved size (FWHM) of each Gaussian component.}
\end{figure*}

Figure 1$a$ shows the map of the 6.9 mm continuum
toward the HH 24 MMS region
from the C- and the D-array data combined.
The peak intensity is 5.66 mJy beam$^{-1}$.
HH 24 MMS was clearly detected and shows an extended structure.

Detailed structures of compact objects were revealed
by imaging with the C-array data only (Fig. 1$b$).
HH 24 MMS was resolved into two objects,
and another source was detected $\sim$3$''$ east of HH 24 MMS.
Their coordinates and peak intensities are listed in Table 1.
Source 1, the brightest one, is
near the 3.6 cm continuum source detected by Reipurth et al. (2002)
and is elongated in the southeast-northwest direction.
In contrast, source 2 is elongated in the northeast-southwest direction.
The separation between sources 1 and 2
is $\sim$0.9$''$ or 360 AU at a distance of 400 pc (Anthony-Twarog 1982).

Source 3 was detected over the detection limit of 0.4 mJy beam$^{-1}$ 
(S/N = 6) in the C-array map (Fig. 1$b$),
and it can also be seen
in the C and D-array map (Fig. 1$a$) as a weak peak.
Source 3 is unresolved.
Since it was not detected in other wavelengths,
the nature of source 3 is unclear.

\subsection {Spectral Energy Distribution and Mass}

In low-mass star forming regions,
continuum emission in the centimeter-millimeter band is usually
either free-free radiation from hot ionized gas
or thermal radiation from cold dust, or a combination of both.
The free-free emission usually
has a small ($\lesssim$1) spectral index,
while the dust emission has a large ($\gtrsim$2) spectral index
(Reynolds 1986; Anglada et al. 1998).
The SED was examined to investigate the emission mechanism of HH 24 MMS.
Since sources 1 and 2 are resolved in the 6.9 mm band only,
SED of the whole HH 24 MMS system was constructed.
The total flux of the HH 24 MMS system,
measured in a 4$''$ $\times$ 4$''$ box,
is 12.0 $\pm$ 0.5 mJy (from the map shown in Fig. 1$a$),
and the SED in the wavelength range of 3.6 cm to 3.4 mm
is shown in Figure 2.

We first tried to fit the SED of HH 24 MMS
as a sum of two power-law components,
one from the free-free emission and the other from the dust emission,
over the wavelength range of 3.4 mm to 3.6 cm (8.3--88 GHz).
The best fit has spectral indices of 2.1 $\pm$ 0.3 and 3.7 $\pm$ 0.2
(Fig. 2 {\it left panel}).
It was found that acceptable fits can only have
large ($>$2) spectral indices for both components,
which clearly means that the emission is mostly from dust
and that the free-free emission is negligible
in the wavelength range considered.

Then, as the dust emission seems to be dominating
even in the centimeter wavelengths,
the SED was fitted again with a single-component power-law distribution.
The fits were made using three data points, excluding the 3.6 cm data,
to avoid any contribution from free-free emission.
The best fit has a spectral index of $\alpha$ = 2.8 $\pm$ 0.2
in the millimeter range (Fig. 2 {\it right panel}).

Extrapolating the best-fit SED to longer wavelengths,
a substantial fraction ($\sim$90\%) of the 3.6 cm flux
seems to come from the dust emission.
Therefore, the free-free emission, if any, must be very weak.
If 10\% of the 3.6 cm flux comes from optically thin free-free emission,
its contribution to the 6.9 mm flux would be only $\sim$0.2\% (or 0.02 mJy).
That is, the 6.9 mm flux of HH 24 MMS is almost entirely from dust.

As the millimeter continuum flux is mostly from dust,
the mass of the emission structure can be estimated from the SED.
To derive the mass from the dust continuum flux,
the mass opacity given by Beckwith \& Sargent (1991) is assumed,
\begin{equation}
   \kappa_\nu = 0.1 \left(\frac{\nu}{\nu_0}\right)^{\beta}
                {\rm cm^2\ g^{-1}},
\end{equation}
where $\nu$ is the frequency, $\nu_0$ = 1200 GHz,
and $\beta$ is the opacity index.
This opacity is for total mass of gas and dust.
The opacity index of HH 24 MMS
is $\beta \approx \alpha - 2 = 0.8$.
This small value of $\beta$ implies the presence of large dust grains
probably caused by the grain growth in the high density environment
(e.g., Miyake \& Nakagawa 1993),
which was pointed out by several authors previously
(Chandler et al. 1995; Ward-Thompson et al. 1995).

The dust temperature, $T_d$, used in previous works
ranges from 12 K to 100 K (Phillips et al. 2001; Chandler et al. 1995).
Here we adopt $T_d$ = 20 $\pm$ 5 K from Ward-Thompson et al. (1995).
In the 6.9 mm maps,
the peak brightness temperature is about 8 K,
and the mean brightness temperature in a 4$''$ $\times$ 4$''$ box
is less than 1 K.
Since the brightness temperature is much lower than the dust temperature,
the optical depth of the 6.9 mm continuum emission must be small.
Assuming optically thin emission, the mass can be estimated by
\begin{equation}
   M = \frac{F_\nu D^2}{\kappa_\nu B_\nu(T_d)},
\end{equation}
where $F_\nu$ is the flux density, $D$ is the distance to the source,
and $B_\nu$ is the Planck function.
Assuming $D$ = 400 pc, the total mass is 1.4$^{+1.5}_{-0.7}$ $M_{\odot}$,
which is consistent with the mass estimate of Chandler et al. (1995).
This mass includes both the accretion disk(s)
and the inner protostellar envelope.

\begin{figure*}
\epsscale{0.9}
\plotone{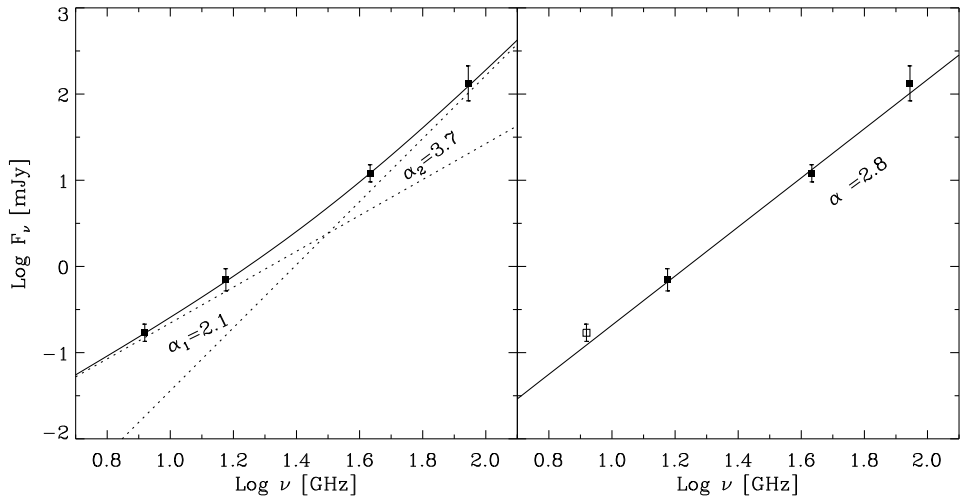}
\caption{
Spectral energy distribution of HH 24 MMS (sources 1 and 2 combined).
Flux densities are from Reipurth et al. (2002),
Chandler et al. (1995), and this work.
{\it Solid curves}:
Best-fit power-law spectra.
{\it Left panel}:
Two-component power-law spectrum.
The best-fit spectral indices are
$\alpha_1 = 2.1 \pm 0.3$ and $\alpha_2 = 3.7 \pm 0.2$.
All the four data points were considered for the fit.
{\it Dotted lines}:
Each component of the fit.
{\it Right panel}:
Single-component power-law spectrum.
The best-fit spectral index is $\alpha = 2.8 \pm 0.2$.
The 8.4 GHz data point was not considered for the fit.}
\end{figure*}

\section{DISCUSSION}

\subsection {Decomposition of the HH 24 MMS System}

At the first glance of the 6.9 mm maps,
the HH 24 MMS system appears
like a combination of a protostar (source 1) and an outflow (source 2).
However, the SED shows that, for both sources 1 and 2,
the 6.9 mm flux is mostly from dust.
Especially, the 6.9 mm flux of source 2 must be entirely from dust
because the 3.6 cm source is not associated with source 2 (Fig. 1$b$).
That is, it is very unlikely that source 2 is a thermal jet.

Since both sources are extended and partially overlapping,
a simple two-component fitting to the image was done
to decompose the system.
The image was fitted with a sum of two elliptical Gaussian distributions.
Only the pixels around source 1 and 2
with intensities above 0.3 mJy beam$^{-1}$ were considered for the fit,
and source 3 was ignored.
Table 1 lists the parameters of the best-fit Gaussian profiles,
and Figure 1$c$ shows the best-fit intensity distribution,
which agrees with the observed image within three times the rms noise.
The total flux of the two-component Gaussian profile
also agrees with that of the real image within 0.1 mJy.

There is a bipolar jet seen in H$_2$ emission in the HH 24 MMS region
with the axis in the northeast-southwest direction
(Bontemps et al. 1996).
The near-IR H$_2$ map of Davis et al. (1997) shows
that HH 24 MMS is almost certainly the driving source of this outflow.
The position angle of these H$_2$ knots (NE 1, NE 2, and SW 2)
with respect to HH 24 MMS is 32\arcdeg\ $\pm$ 6\arcdeg,
which is perpendicular to the elongation of source 1.
Therefore, source 1 seems to be the driving source of the bipolar outflow.

The peak position of the 3.6 cm source (Reipurth et al. 2002)
does not exactly coincide with the position of source 1 (Fig. 1$b$).
The position displacement is about 0.7$''$.
Since the 3.6 cm source is weak and extended ($\sim$1$''$),
it is not clear that the displacement is significant.
In this paper we assume
that the 3.6 cm source is identical to or closely associated with source 1,
but we cannot rule out the possibility
that the 3.6 cm source indicates
the existence of yet another young stellar object.
This issue should be resolved by more sensitive observations
of the centimeter continuum in the future.

For source 2, three kinds of interpretation are possible. 
(1) Source 2 is another protostar
    with a projected separation of 360 AU from source 1.
(2) Source 2 is a clump of dust swept and heated by the outflow.
(3) Source 2 has multiple clumps in the dense envelope as seen in
    other protostars such as NGC 1333 IRAS 4B (Looney et al. 2000).
The first possibility seems to be the simplest explanation.
The second possibility is rather unlikely
because the position angle of the elongation of source 2 ($\sim$69\arcdeg)
is different from the direction of the outflow by $\sim$40\arcdeg.
In addition, the mass of source 2 seems too large for a clump around outflow.
The third possibility can only be tested by imaging with a higher resolution,
but it is rather peculiar
that only one side of the envelope has dense clumps.
Therefore, we prefer the interpretation
that HH 24 MMS may be a protostellar binary system,
and we will adopt this interpretation in the following discussions.

The separation between sources 1 and 2 is
within the range of the expected separation of young binaries.
In a few examples of well-studied Class 0 protostars,
observations and modeling suggest
that the centrifugal radius ranges from 34 AU to 600 AU
(see Myers et al. 2000 and references therein).
Reipurth et al. (2004) carried out a radio-continuum survey
of embedded objects with distances between 140 and 800 pc
and found many binaries with separations ranging from 0.5$''$ to 12$''$.
Many numerical simulations of core fragmentation
produce multiple systems with separations of a few hundreds of AU
(Goodwin et al. 2007).

Both components of HH 24 MMS are highly elongated,
with axial ratios of 3 for source 1 and 8 for source 2,
which suggests that each of them may be
a disk-like flattened structure viewed nearly edge-on.
The size of each component is somewhat larger
than the expected size of an accretion disk
supported by Keplerian rotation ($\sim$100 AU),
and each of the 6.9 mm continuum sources may consist of
a compact accretion disk
surrounded by a flattened inner envelope or a pseudodisk
(a contracting disk-like structure
that is not completely supported by rotation;
Galli \& Shu 1993; Tomisaka 2002; Allen et al. 2003).
Assuming that the dust properties of each source
are not very different from the overall dust properties
of the HH 24 MMS system,
the mass can be derived from the flux of each source (Table 1).

\subsection {Misaligned Disks}

The majority of stars belong to multiple-star systems
(Duquennoy \& Mayor 1991),
and understanding of star formation cannot be complete
without knowing the physical processes in multiple-protostar systems
(Tohline 2002).
Study of the protobinary in the earliest evolutionary stage
is essential in understanding how the fragmentation occurs
and what is the initial condition of the binary evolution.
There are two main mechanisms for fragmentation:
(1) rotational fragmentation and (2) turbulent fragmentation
(Goodwin et al. 2007).
The outcome of these two mechanisms can be quite different.
The former occurs
when the kinematics of the initial core is dominated by systematic rotation,
and the resulting binary would be well-ordered,
i.e., disks would be well-aligned.
In contrast, the latter occurs
when the core is dominated by random motions,
and the resulting binary is likely to have misaligned disks.
Therefore, the alignment of disks is one of the key observational measures
that can provide strong constraints on the fragmentation mechanism.

In principle, multiple outflows
can give some implications on the degree of binary disk alignment
because outflows are ejected perpendicular to the circumstellar disks.
There are some examples of misaligned systems inferred from jets 
emanating with different position angles 
(Gredel \& Reipurth 1993; Reipurth et al. 1993). 
However, this method is problematic because two well-aligned outflows
would be misidentified as a single outflow.
The best way to test the disk alignment
may be the direct imaging of both disks.

If HH 24 MMS is a binary system as discussed in the previous section,
it is one of the rare examples of protobinary system
imaged with an angular resolution high enough
to investigate the degree of alignment.
Each source may consist of a compact accretion disk
surrounded by a pseudodisk,
and the axis of the accretion disk is very likely
to be parallel to the axis of the pseudodisk around it.
Then the axis direction of the accretion disk can be inferred
from the minor axis of the elliptical fit to the 6.9 mm source.

The major axes of the two components of the HH 24 MMS system
are not aligned at all.
The difference in the position angle between the binary components, 
from the Gaussian fit (Table 1), 
is $\sim$45\arcdeg\ (or $\sim$135\arcdeg\ if they are retrograde).
HH 24 MMS makes a stark contrast
to the other known example of resolved binary disks, L1551 IRS 5.
In the L1551 IRS 5 system, the two disks are very well aligned
(Rodr{\'\i}guez et al. 1998).
The difference between the two systems
is probably owing to the binary separation:
360 AU in HH 24 MMS and 45 AU in L1551 IRS 5.
In the classification scheme of multiple young stellar objects
suggested by Looney et al. (2000),
HH 24 MMS belongs to the common-envelope type
while L1551 IRS 5 belongs to the common-disk type.
We speculate that the rotational fragmentation
is important in close binaries
and the turbulent fragmentation is important in wide binaries,
though more examples are certainly needed
to draw a statistically significant conclusion.
Other examples may include IRAS 16293--2422 and 
NGC 1333 IRAS 4A (Looney et al. 2000),
but their structures are complicated
and their disk orientations are less obvious.

Though studies through direct imaging is difficult,
indirect evidences show that misaligned systems are not necessarily unusual.
High-resolution near-IR polarimetry revealed
plenty of examples of misaligned binaries in the Class II phase
(Monin et al. 2007).
Disks in T Tauri binaries are not perfectly coplanar 
(Jensen et al. 2004),
and the relative alignment of the disks may decrease after formation 
(Bate et al. 2000).
Therefore, HH 24 MMS can be considered
as a {\it precursor} of such misaligned T Tauri systems.
As HH 24 MMS is a rare example of a wide binary
that can be imaged well enough to study the disk alignment,
this system gives us a chance to examine 
the initial conditions of binary evolution. 
 
\acknowledgments

We thank K.-T. Kim for helpful discussions and encouragement.
M. K. is grateful to J. Bieging for helpful discussions.
This work was partially supported by the LRG program of KASI.
M. K. acknowledges support from the Korea Research Foundation 
grant KRF-2007-612-C00050 funded by the Korean Government (MOEHRD).

\end{document}